\documentclass[aps,prd,amsfonts,amssymb,amsmath,showpacs,twocolumn]{revtex4}

\usepackage{graphicx}
\usepackage{dcolumn}
\usepackage{bm}

\begin{document}

\title{Instability of a four-dimensional de Sitter black hole with a
conformally coupled scalar field}

\author{Tom J.T. Harper}
\author{Paul A. Thomas}
\author{Elizabeth Winstanley}
 \email{E.Winstanley@shef.ac.uk}
\author{Phil M. Young}
\affiliation{Department of Applied Mathematics, The University of Sheffield,
Hicks Building, Hounsfield Road, Sheffield, S3 7RH, United Kingdom.}

\date{\today}

\begin{abstract}
We study the stability of new neutral and electrically charged
four-dimensional black hole solutions of Einstein's equations with
a positive cosmological constant and conformally coupled scalar field \cite{mtz}.
The neutral black holes are always unstable.
The charged black holes are also shown analytically to be unstable for
the vast majority of the parameter space of solutions,
and we argue using numerical techniques
that the configurations corresponding to the remainder of the parameter space are
also unstable.
\end{abstract}

\pacs{04.20.Jb, 04.40.Nr, 04.70.Bw}

\maketitle

\section{Introduction}
\label{sec:intro}

Black hole solutions of the four-dimensional Einstein-scalar field system have been extensively
studied for over thirty years, with a particular focus on
proving uniqueness (``no-hair'') theorems in various models
(see, for example, \cite{heusler} for a review).
Within these models, conformal coupling of the scalar field
is of particular interest,  both with and without
an additional coupling to the Maxwell field
(see, for example, \cite{ew} for a
brief review of work on this situation).
In asymptotically flat space, with no scalar self-interaction potential,
there is an exact, closed form solution (the BBMB solution) \cite{BBM,bek1,bek2}, which
has not been without controversy \cite{sudarsky} because the scalar field diverges on the
event horizon.
Furthermore, it is known that this solution is unstable \cite{bronnikov}.
For space-times which are asymptotically anti-de Sitter,
with zero potential or a quadratic scalar field potential, numerical solutions exist in four
dimensions \cite{ew}, of which at least some are linearly stable
under spherically-symmetric perturbations.
Interestingly, the corresponding three-dimensional black hole solution is known
in closed form \cite{mz}, but is unstable \cite{martinez}.

Asymptotically de Sitter geometries are the focus of this paper.
For minimally coupled scalar fields, non-trivial scalar field hair is
possible if the scalar field potential is non-convex \cite{cai,torii},
although the hair is unstable \cite{torii}.
For conformally coupled scalar fields,
if the scalar field potential is zero or quadratic, then there are no
non-trivial black hole solutions \cite{ew}.
However, in the presence of a quartic self-interaction potential
there is an exact, closed-form solution for both charged and neutral black holes
found recently by Martinez, Troncoso and Zanelli \cite{mtz}, which
we shall refer to hereafter as the MTZ solution.
This solution is the de Sitter analogue of the BBMB solution,
although the scalar field is regular on and outside the event horizon.
The purpose of this paper is to discover whether, like the BBMB solution,
the MTZ black hole is unstable.

The outline of this paper is as follows.
In section \ref{sec:mtz} we briefly review the neutral and electrically charged
MTZ solutions \cite{mtz},
whose stability is then studied using linear perturbation theory
in sections \ref{sec:pertneut} and \ref{sec:pertcharged}
respectively.
In the appendix we outline the proof of a result needed
for the stability analysis of the charged solutions.
Finally, we present our conclusions in section \ref{sec:conc}.
The metric has signature $(-+++)$ and we use units in which
$c=8\pi G=1$ throughout.

\section{The MTZ solution}
\label{sec:mtz}

We begin with the action for gravity with a conformally coupled scalar field with a quartic
self-interaction potential and an electromagnetic field \cite{mtz}:
\begin{eqnarray}
S & = &
\frac{1}{2}
\int d^{4}x \sqrt{-g}
\left[
R-2\Lambda
-g^{\mu\nu} \partial_{\mu} \phi \partial_{\nu} \phi
-\frac{1}{6}R\phi^{2}
\right.
\nonumber \\
& &
\left. \qquad \qquad
- 2\alpha\phi^{4}
-\frac{1}{8\pi}F^{\mu\nu}F_{\mu\nu}
\right] ;
\label{eq:action}
\end{eqnarray}
where $\alpha $ is the coupling constant.
For neutral black holes, the electromagnetic field is absent.
The Einstein and scalar field equations derived from this action are
\begin{subequations}
\begin{eqnarray}
G_{\mu\nu}+\Lambda g_{\mu\nu}-
T_{\mu\nu}^{\phi}-T_{\mu\nu}^{EM} & = & 0 ;
\label{eq:eine}
\\
\Box\phi-\frac{1}{6}R\phi-4\alpha\phi^{3}& = & 0 ;
\label{eq:sca}
\end{eqnarray}
\end{subequations}
where $\Box \equiv g^{\mu\nu}\nabla_{\mu}\nabla_{\nu}$, and the
scalar and electromagnetic stress energy tensors are
\begin{eqnarray*}
T_{\mu\nu}^{\phi} & = &
\partial_{\mu} \phi \partial_{\nu} \phi
-\frac{1}{2} g_{\mu\nu} g^{\alpha\beta}
\partial_{\alpha} \partial_{\beta} \phi
\\
& &
+\frac{1}{6}\left[ g_{\mu\nu}\Box-\nabla_{\mu}\nabla_{\nu}
+G_{\mu\nu} \right] \phi^{2}
-\alpha g_{\mu\nu}\phi^{4};
\\
T_{\mu\nu}^{EM} & = &
\frac{1}{4\pi} \left( g^{\alpha\beta}F_{\mu\alpha}F_{\nu\beta} -
\frac{1}{4} g_{\mu\nu} F_{\alpha\beta}F^{\alpha\beta} \right) .
\end{eqnarray*}
For charged black holes, we also have the Maxwell equations
\begin{equation}
\nabla_{\mu}F^{\mu\nu}  =  0; \qquad \qquad
F_{[\mu\nu;\lambda]}  =  0.
\label{eq:em}
\end{equation}

For both the charged and electrically neutral cases,
the field equations are solved by the spherically symmetric metric
\begin{equation}
\label{eq:met}
ds^{2}=-N(r)dt^{2}+ N(r)^{-1}dr^{2}+
r^{2}\left( d\theta^{2}+\sin^{2}\theta \, d\varphi^{2}\right) ;
\end{equation}
where
\begin{equation*}
N(r)=-\frac {\Lambda}{3} r^{2}+
\left(1-\frac{M}{r}\right)^{2} ;
\end{equation*}
and the scalar curvature is a constant given by
\begin{equation*}
R=4\Lambda .
\end{equation*}
The geometry (\ref{eq:met}) is that of the Reissner-Nordstr\"om-de Sitter
black hole with inner, event and cosmological horizons at values of the
radial co-ordinate $r$ given by, respectively,
\begin{eqnarray}
r_{-} & = &
\frac {l}{2} \left[
-1 + {\sqrt {1+ \frac {4M}{l}}}
\right] ;
\nonumber \\
r_{+} & = &
\frac {l}{2} \left[
1 - {\sqrt {1 - \frac {4M}{l}}}
\right] ;
\nonumber \\
r_{++} & = &
\frac {l}{2} \left[
1 + {\sqrt {1 - \frac {4M}{l}}}
\right] ;
\label{eq:hor}
\end{eqnarray}
where $l={\sqrt {3/\Lambda }}$.
From (\ref{eq:hor}), it is clear that the solution is defined only for $0<M<M_{max}=l/4$.

The form of the scalar field is different for the electrically neutral and charged models.
In the case with no electromagnetic field, there is a solution only if the coupling
constant $\alpha $ is given by
\begin{equation*}
\alpha = -\frac {1}{36} \Lambda ;
\end{equation*}
and then the scalar field takes the form
\begin{equation}
\phi (r) = \frac {{\sqrt {6}}M}{r-M} ;
\label{eq:phineut}
\end{equation}
which has a pole at $r=M<r_{+}$, lying inside the event horizon.

For charged black holes,
the only non-vanishing component of the electromagnetic field is
\begin{equation*}
F_{tr}=-\partial_{r}A_{t}=\frac{Q}{r^{2}};
\end{equation*}
where the charge-to-mass ratio is given by
\begin{equation}
\label{eq:qm}
\left(\frac{Q}{M}\right)^{2}
=8\pi\left(1+\frac{\Lambda }{36\alpha}\right) ;
\end{equation}
and the scalar field in this case is
\begin{equation}
\label{eq:scc}
\phi(r)=\sqrt{-\frac{\Lambda}{6\alpha}}\frac{M}{r-M} .
\end{equation}
This latter solution only exists
provided $\alpha$ satisfies the bound
\begin{equation*}
36\alpha<-\Lambda .
\end{equation*}

\section{Instability of the MTZ solution: Neutral case}
\label{sec:pertneut}

We now analyze the stability of the MTZ black holes using linear
perturbation theory.
The electrically neutral and charged solutions need to be considered
separately, and we begin with the neutral case
since this is the simpler.
In this and the following section, we consider only spherically
symmetric perturbations since these are sufficient
to show instability.
Many calculations were performed using the GRTensorII \cite{grtensor}
routine {\tt {linpert}} \cite{linpert}.

The perturbed spherically symmetric metric takes the form:
\begin{eqnarray}
ds^{2} & = &  -N(t,r)
e^{2\varepsilon\hat{\delta}(t,r)}dt^{2}
+N^{-1}(t,r)dr^{2}
\nonumber \\ & &
+r^{2}
(d\theta^{2}+\sin^{2}\theta\,d\varphi^{2});
\label{eq:pmet}
\end{eqnarray}
where
\begin{equation}
\label{eq:N}
N(t,r)=N(r)+\varepsilon\hat{N}(t,r);
\end{equation}
and $\varepsilon $ is a small parameter.
The scalar field is perturbed as
\begin{equation}
\label{eq:pphi}
\phi(t,r)= \phi(r)+\varepsilon\hat{\phi}(t,r).
\end{equation}
The ansatz (\ref{eq:pmet}-\ref{eq:pphi}) can then be substituted into
the Einstein and scalar field equations (\ref{eq:eine}) and (\ref{eq:sca}),
the equations then linearized and the metric perturbations ${\hat {N}}$ and ${\hat {\delta }}$
eliminated to yield a single perturbation equation for the perturbed scalar
field ${\hat {\phi }}$.
While this is possible in principle, in practice the algebra becomes somewhat
unwieldy, even using a computer algebra package.

We therefore employ a transformation to a simpler system which makes the
calculations more tractable, and then substitute back the original
perturbations.
Under the following conformal transformation \cite{maeda}
(see also \cite{ew} for more details),
\begin{equation}
\tilde{g}_{\mu\nu}=\Omega^{2}g_{\mu\nu};
\label{eq:cmet}\\
\end{equation}
where
\begin{equation*}
\Omega=
\left( 1- \frac {1}{6}\phi^{2} \right) ^{\frac{1}{2}} ;
\end{equation*}
the action (\ref{eq:action}) becomes that of a minimally coupled scalar field $\Phi $
with potential $V(\Phi )$, with
\begin{equation}
\Phi=\sqrt{6}\tanh^{-1}\left(\frac{\phi}{\sqrt{6}}\right) ;
\label{eq:cphi}
\end{equation}
and
\begin{equation*}
V(\Phi )= 2\Lambda \sinh ^{2} \left( \frac {\Phi }{{\sqrt {6}}} \right) .
\end{equation*}
At this stage we should comment that,
from Eq. (\ref{eq:phineut}), there is always a point
($r=2M$) between the event and cosmological
horizons at which $1-\frac {1}{6} \phi ^{2}=0$, and therefore the conformal transformation
(\ref{eq:cmet}) breaks down at this point.
However, since this is only a single point, and not an open set, this does not render
the theory completely ill-defined (see, for example, the discussion in Ref. \cite{ashtekar}).
In addition, the transformed variables have no physical significance;
the only purpose of the transformation is to simplify the algebra, in particular
to derive the master perturbation equation, as in the analysis of Ref. \cite{bronnikov}.
Our conclusions will be drawn only once the perturbations of the original system have been
substituted.

The new metric (\ref{eq:cmet}) takes the form \cite{ew}:
\begin{equation}
\label{eq:cmete}
d\tilde{s}^{2}=-{\mathcal{N}}e^{2\Delta}dt^{2}+{\mathcal{N}}^{-1}
dx^{2}+x^{2}(d\theta^{2}+\sin^{2}\theta \, d\varphi^{2}) ;
\end{equation}
where we
have transformed the radial coordinate by
\begin{equation*}
x=\left(1-\frac{1}{6}\phi^{2}\right)^{\frac{1}{2}}r ;
\end{equation*}
and the metric quantities $\mathcal{N}$ and
$\Delta$ are related to the original metric variables by \cite{ew}:
\begin{eqnarray*}
\mathcal{N} & = &
N\left(1-\frac{1}{6}\phi^{2}-\frac{1}{6}r\phi\phi'\right)^{2}
\left(1-\frac{1}{6}\phi^{2}\right)^{-2} ;
\\
e^{\Delta} & = &
\left(1-\frac{1}{6}\phi^{2}-\frac{1}{6}r\phi\phi'\right)^{-1}
\left(1-\frac{1}{6}\phi^{2}\right)^{\frac{3}{2}} .
\end{eqnarray*}
Note that $\Delta \neq 0$ in the new metric.
The transformed variables ${\mathcal {N}}$, $\Delta $ and $\Phi $
are perturbed as follows:
\begin{eqnarray*}
\mathcal{N}(t,x) & = & \mathcal{N}(x)+\varepsilon\hat{\mathcal{N}}(t,x) ;\\
\Delta(t,x) & = &
\Delta(x)+\varepsilon\hat{\Delta}(t,x) ; \\
\Phi(t,x) & = & \Phi(x)+\varepsilon\hat{\Phi}(t,x) .
\end{eqnarray*}
These expressions are substituted into the new Einstein and scalar field equations
(a tilde denotes quantities calculated using the transformed metric (\ref{eq:cmete})):
\begin{eqnarray*}
0 & = &
\tilde{G}_{\mu\nu}+\Lambda \tilde{g}_{\mu\nu}
- \tilde{\nabla}_{\mu}\Phi\tilde{\nabla}_{\nu}\Phi
+\frac{1}{2}\tilde{g}_{\mu\nu}(\tilde{\nabla}\Phi)^{2}
\\ & &
+\tilde{g}_{\mu\nu}
V(\Phi) ;
\\
0 & = &
{\tilde{\Box}}\Phi-\frac{\partial V}{\partial\Phi} ;
\end{eqnarray*}
only terms linear in the parameter $\varepsilon $ are retained,
and the metric perturbations can then be eliminated.
A single perturbation equation is then found
for the variable ${\hat {\phi }}$.
For periodic perturbations ${\hat {\phi }} (t,x) = e^{i\sigma t} {\hat {\phi }} (x)$,
this equation takes the form:
\begin{equation}
-\frac {\partial ^{2}{\hat {\phi }}}{\partial x_{*}^{2}}
-{\mathcal {Q}}_{1} \frac {\partial {\hat {\phi }}}{\partial x_{*}}
-{\mathcal {Q}}_{2} {\hat {\phi }} + {\mathcal {U}} {\hat {\phi }}
=\sigma ^{2} {\hat {\phi }} ;
\label{eq:phiperteqn}
\end{equation}
where we have introduced the usual ``tortoise'' co-ordinate $x_{*}$ defined by
\begin{equation}
\frac {dx_{*}}{dx} = \frac {1}{{\mathcal {N}}e^{\Delta }};
\label{eq:tort1}
\end{equation}
so that
\begin{equation}
\frac {dx_{*}}{dr} = \frac {1}{N};
\label{eq:tort2}
\end{equation}
and the region between the event and cosmological horizons $r_{+}<r<r_{++}$ becomes
$-\infty < x_{*} < \infty $.
The quantities ${\mathcal {Q}}_{1}$ and ${\mathcal {Q}}_{2}$ are given by
\begin{eqnarray*}
{\mathcal {Q}}_{1} & = &
\frac {2N}{r} {\mathcal {A}}^{-1} {\mathcal {D}} ;
\\
{\cal {Q}}_{2} & = &
\frac {N}{r} \left[
\frac {dN}{dr} {\mathcal {A}}^{-1} {\mathcal {D}}
+ \frac {1}{2} N {\mathcal {A}}^{-2} {\mathcal {D}} \phi \frac {d\phi }{dr}
+ N {\mathcal {A}}^{-1} \frac {d{\mathcal {D}}}{dr} \right] ;
\end{eqnarray*}
and the perturbation potential ${\mathcal {U}}$ takes the form
\begin{eqnarray}
{\mathcal {U}} & = &
\frac {N}{r^{2}} \left[
1 - N {\mathcal {A}}^{-2} {\mathcal {B}}^{2}
- r^{2} \left( \frac{d\phi }{dr} \right) ^{2}
{\mathcal {B}}^{-2}
\right.
\nonumber \\
& &
+\frac {4}{3} \Lambda r^{3} \phi \frac {d\phi }{dr}
{\mathcal {B}}^{-1}
+ \Lambda r^{4} \left( \frac {d\phi }{dr} \right) ^{2}
{\mathcal {B}}^{-2} \left( 1 + \frac {1}{6} \phi ^{2} \right)
\nonumber \\
& & \left.
- \frac {1}{3} \Lambda r^{2} \left( 1 + \frac {1}{6} \phi ^{2} \right)
\right] ;
\label{eq:neutpot}
\end{eqnarray}
where
\begin{eqnarray}
{\mathcal {A}} & = &
1 - \frac {1}{6} \phi ^{2} ;
\nonumber
 \\
 {\mathcal {B}} & = &
 1 -\frac {1}{6} \phi ^{2} - \frac {1}{6} r \phi \frac {d\phi }{dr} ;
 \nonumber
 \\
 {\mathcal {D}} & = &
 1 - \frac {1}{6} \phi ^{2} + \frac {1}{6} r \phi \frac {d\phi }{dr} .
 \label{eq:AandB}
\end{eqnarray}
It should be stressed that the equation (\ref{eq:phiperteqn}) could equally well have been
derived directly, without using the conformal transformation (\ref{eq:cmet}),
but that our method has simplified the algebra considerably.
Furthermore, the equation (\ref{eq:phiperteqn}) holds everywhere, although the conformal
transformation (\ref{eq:cmet}) is defined only for those values of $r$ for which $\phi ^{2}<6$.

The perturbation equation (\ref{eq:phiperteqn}) has a regular
singular point at $r=r_{0}=2M$, where $\phi ^{2}=6$.
Using the standard Frobenius method, near this point ${\hat {\phi }}$ behaves like
\begin{equation}
{\hat {\phi }} \sim \left( r - r_{0} \right) \left[
C_{1} + C_{2} \log \left( r - r_{0} \right)
\right]
\label{eq:phibehav}
\end{equation}
for constants $C_{1}$ and $C_{2}$.
The presence of the regular singular point means that it is not possible to define a non-singular
transformation of (\ref{eq:phiperteqn}) to standard Schr\"odinger form, as, for example, was
possible for three-dimensional black holes in Ref. \cite{martinez}.

\begin{figure}
\includegraphics[width=6cm,angle=270]{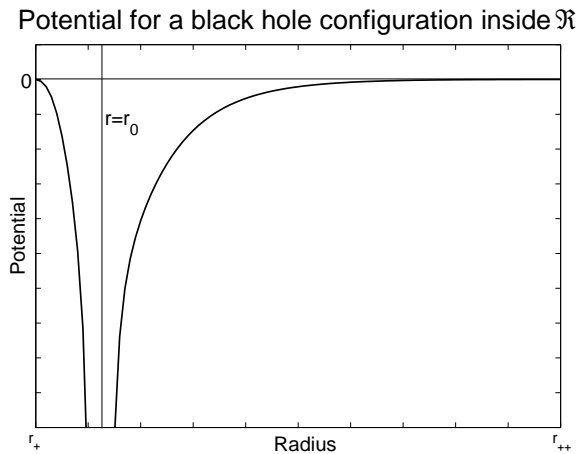}
\caption{Form of the potential ${\mathcal {U}}$ for the neutral
MTZ black holes. The potential has a double pole at $r=r_{0}=2M$.
This is also the form of the potential for the charged MTZ black holes
configurations corresponding to points within the region $\Re $, see
section \ref{sec:pertcharged}.}
\label{fig:pot1}
\end{figure}

In order to cast the perturbation equation (\ref{eq:phiperteqn}) into standard Schr\"odinger form,
we define a new perturbation variable $\Psi $ as follows \cite{bronnikov,martinez}:
\begin{equation}
\Psi = r \left| 1- \frac {1}{6} \phi ^{2} \right| ^{-\frac {1}{2}}
{\hat {\phi }} ,
\label{eq:Psi}
\end{equation}
so that the equation (\ref{eq:phiperteqn}) then takes the form
\begin{equation}
-\frac {\partial ^{2}\Psi }{\partial x_{*}^{2}} + {\mathcal {U}} \Psi
=\sigma ^{2} \Psi ;
\label{eq:Psischro}
\end{equation}
with potential ${\mathcal {U}}$ given by (\ref{eq:neutpot}).
It is straightforward to show that the quantity ${\mathcal {B}}$ (\ref{eq:AandB})
is regular and non-zero everywhere between the event and cosmological horizons.
Therefore, the potential (\ref{eq:neutpot}) is regular for all $x_{*}$ apart from the
second term, which has a double pole at $r=r_{0}=2M$ (where $\phi ^{2}=6$).
Near $r=r_{0}=2M$, the potential has the behaviour
\begin{equation*}
{\mathcal {U}} = -\frac {1}{4} \left[ N(r_{0})\right] ^{2} \left( r-r_{0} \right) ^{-2} +
O\left( r-r_{0}\right) ^{-1};
\end{equation*}
and ${\mathcal {U}} \rightarrow 0 $ as $x_{*} \rightarrow \pm \infty $.
The form of the potential is sketched in Fig. \ref{fig:pot1}.
Like the original equation (\ref{eq:phiperteqn}), the new perturbation equation
(\ref{eq:Psischro}) has a regular singular point at $r=r_{0}$, and a Frobenius
expansion about this point gives the behaviour of $\Psi $ to be
\begin{equation*}
\Psi \sim \left( r - r_{0} \right) ^{\frac {1}{2}} \left[
{\tilde {C}}_{1} + {\tilde {C}}_{2} \log \left( r - r_{0} \right) \right] ,
\end{equation*}
where ${\tilde {C}}_{1}$ and ${\tilde {C}}_{2}$ are constants.
Using equation (\ref{eq:phibehav}), it can be seen that the transformation
(\ref{eq:Psi}) yields precisely the correct behaviour for $\Psi $, so that the
two perturbation equations (\ref{eq:phiperteqn}) and (\ref{eq:Psischro}) are
equivalent, in particular they have the same number of negative eigenvalues.

Using a standard result in quantum mechanics \cite{bronnikov}, potentials of this form have
an infinite number of bound states corresponding to $\sigma ^{2}<0$, that is,
an infinite number of unstable modes.
Therefore we also have an infinite number of unstable modes of the original perturbations.
This is the same behaviour as found for the BBMB black hole~\cite{bronnikov}.

\section{Instability of the MTZ solution: Charged case}
\label{sec:pertcharged}

The perturbation analysis of the electrically charged black holes is considerably
more complicated, and although we shall again make use of the conformal transformation
(\ref{eq:cmet}) to simplify the algebra, the untransformed perturbations
themselves can be simplified first.

Making a choice of gauge to set $A_{r}=0$, the gauge
potential giving rise to the electromagnetic field is perturbed as
below:
\begin{eqnarray*}
A_{t} & = & \frac{Q}{r}+\varepsilon\hat{A}_{t}(t,r) ;
\\
A_{r} & = & 0 ;
\\
A_{\theta} & = & \varepsilon\hat{A}_{\theta}(t,r) ;
\\
A_{\varphi} & = &
\varepsilon\hat{A}_{\varphi}(t,r)\sin\theta ;
\end{eqnarray*}
so that the second of the Maxwell equations (\ref{eq:em}) is
automatically satisfied.
The metric and scalar field are perturbed in the same way as for the
neutral black holes (\ref{eq:pmet}--\ref{eq:pphi}).
The $(t \theta )$, $(r\theta)$, $(t \varphi )$ and
$(r\varphi )$ components of the Einstein field equation
(\ref{eq:eine}) give
\begin{equation*}
\frac{\partial\hat{A}_{\theta}(t,r)}{\partial
r}=\frac{\partial\hat{A}_{\theta}(t,r)}{\partial
t}=\frac{\partial\hat{A}_{\varphi}(t,r)}{\partial
r}=\frac{\partial\hat{A}_{\varphi}(t,r)}{\partial t}=0 ;
\end{equation*}
and since we can remove an arbitrary constant from the gauge
potential without changing the resulting electromagnetic field we
can, without loss of generality, assume that
\begin{equation*}
\hat{A}_{\theta}=\hat{A}_{\varphi}=0 .
\end{equation*}
\par
The $t$ and $r$ components of the first Maxwell equation (\ref{eq:em})
then take the form:
\begin{eqnarray*}
0 & = &
\frac{\partial}{\partial
r}\hat{\delta}(t,r)Q+r^{2}\frac{\partial^{2}}{\partial
r^{2}}\hat{A}_{t}(t,r)+2r\frac{\partial}{\partial
r}\hat{A}_{t}(t,r)
 ;
\\
0 & = &
\frac{\partial^{2}}{\partial t\partial
r}\hat{A}_{t}(t,r)r^{2}+\frac{\partial}{\partial t}
\hat{\delta}(t,r)Q
.
\end{eqnarray*}
These two equations are compatible and can be integrated to give
\begin{equation*}
\hat{\delta}(t,r)=-\frac{r^{2}}{Q}\frac{\partial}{\partial
r}\hat{A}_{t}(t,r) ;
\end{equation*}
where we have absorbed the arbitrary constant
of integration into the time co-ordinate.
With the form of the electromagnetic potential now fixed, the $(tr)$ component
of the Einstein equations can readily be integrated to yield
\begin{eqnarray}
\hat{N} & = &
r \left[
-\frac{2}{3}N \frac {d\phi }{dr} \hat{\phi}
-\frac{1}{6} \frac {dN}{dr} \phi \hat{\phi}
+\frac{1}{3}N\phi\hat{\phi}+\mathcal{F}(r)
\right]
\nonumber \\ & &
\times
\left(1-\frac{1}{6}\phi^{2}-\frac{1}{6}r\phi\phi'\right) ^{-1};
\label{eq:pnn}
\end{eqnarray}
where ${\mathcal{F}}$ is an arbitrary function of $r$.
Incidentally, equation (\ref{eq:pnn}) also holds for the neutral black holes.

At this point, the perturbation equations are again too complicated to readily be simplified
and we employ the conformal transformation as in the previous section.
As in section \ref{sec:pertneut},
we regard the transformation as a convenient algebraic tool with no physical significance
\cite{bronnikov}.

The transformations of the metric and scalar field are the same as before (\ref{eq:cmet}--\ref{eq:cphi}),
but now the electromagnetic field also transforms \cite{waldgr}:
\begin{equation*}
\tilde{F}_{\mu\nu}=F_{\mu\nu}; \qquad \qquad
{\tilde {F}}^{\mu \nu } =\Omega ^{-4} F^{\mu \nu} ;
\end{equation*}
leaving the electromagnetic stress-energy tensor invariant.
The transformed metric variables and scalar field are perturbed in the same way as in
the previous section.
The only non-zero component of the transformed electromagnetic field is ${\tilde {F}}_{tx}$,
which we write as
\begin{equation*}
{\tilde {F}}_{tx} (t,x) = F(x) + \varepsilon {\hat {F}}(t,x);
\end{equation*}
where
\begin{equation*}
\label{eq:Ftr}
F(x) =\frac{Q}{x^{2}}e^{\Delta} .
\end{equation*}

The electromagnetic and Einstein equations can be used to find the following
simple expressions for the perturbations of the transformed metric:
\begin{eqnarray}
\hat{\mathcal{N}}
& = & -x\mathcal{N}\frac{d \Phi}{dx}\hat{\Phi}+\mathcal{S}(x) ;
\nonumber \\
\hat{\Delta}& = & \frac{\hat{F}}{F} ;
\label{eq:metrels}
\end{eqnarray}
where $\mathcal{S}$ is an arbitrary function of $x$.
With these expressions,
the $(tt)$ component of the linearized Einstein equation gives an
equation of the form
\begin{equation}
\label{eq:eq1}
\mathcal{X}\hat{F}+\mathcal{Y}\hat{\Phi}+\mathcal{G}\mathcal{S}(x)+
\mathcal{H}\frac{\partial\mathcal{S}(x)}{\partial x}=0 ;
\end{equation}
where ${\mathcal {X}}$, ${\mathcal {Y}}$, ${\mathcal {G}}$ and ${\mathcal{H}}$
are functions of $x$ only.
Using the equilibrium field equations, it can be shown that
$\mathcal{X}$ vanishes, and
\begin{equation}
\mathcal{Y}=-2\frac{\partial\Delta}{\partial
x}+x\left(\frac{\partial\Phi}{\partial x}\right)^{2}.
\label{eq:res5}
\end{equation}
Furthermore, it can be shown that $\mathcal{Y}=0$
by writing the right-hand-side of (\ref{eq:res5}) in terms of
the original variables and radial coordinate  $r$.
Equation (\ref{eq:eq1}) can then be integrated to give
\begin{equation*}
\mathcal{S}=K \exp \left( -\int\frac{\mathcal{G}}{\mathcal{H}}dx \right)
=\frac{K}{e^{\Delta}x} ;
\end{equation*}
where $K$ is an arbitrary constant.
Now, using the boundary conditions that $\hat{F}=\hat{\Phi}=0$
at the event horizon, it must be the case that $K=0$
and so $\mathcal{S}(x)=0$ identically.
Using this result, the $(xx)$ component of
the linearized Einstein equations reduces to
a relationship between the electromagnetic field and scalar field perturbations:
\begin{equation}
\label{eq:res6}
\hat{F}\frac{\partial F}{\partial x}
-\frac{\partial\hat{F}}{\partial x}F
= x F^{2}
\frac{\partial\Phi}{\partial x}\frac{\partial\hat{\Phi}}{\partial x} .
\end{equation}

Relations (\ref{eq:metrels},\ref{eq:res6}) are sufficient to eliminate the metric
and electromagnetic perturbations from the linearized scalar field equation.
Using the ``tortoise'' co-ordinate (\ref{eq:tort1}--\ref{eq:tort2}),
and following the method of section \ref{sec:pertneut},
we finally arrive at the Schr\"odinger-like equation
\begin{equation}
\label{eq:sch}
-\frac{\partial^{2}\Psi}{\partial x_{*}^{2}}+\mathcal{C}\Psi
=\sigma^{2} \Psi ;
\end{equation}
where $\Psi $ is given by (\ref{eq:Psi}) and
we are once again considering periodic perturbations.
As might be expected, the potential ${\mathcal {C}}$ is more complicated
than for the neutral black holes:
\begin{eqnarray}
{\mathcal {C}} & = &
\frac {N}{r^{2}} \left[
1 - N {\mathcal {A}} ^{-2} {\mathcal {B}}^{2}
 -2 \Lambda r^{2} - 48 \alpha r^{2} - 12 \alpha r^{2} {\mathcal {A}}
\right.
\nonumber \\
& &
+60 \alpha r^{2} {\mathcal {A}}^{-1}
+ \frac {5}{3} \Lambda r^{2} {\mathcal {A}} ^{-1}
- \frac {1}{r^{2}} M^{2} {\mathcal {A}}^{-1}
\nonumber \\
& &
-\frac {1}{36r^{2}}\frac {\Lambda }{\alpha } M^{2} {\mathcal {A}}^{-1}
-r^{2} \left( \frac {d\phi }{dr} \right) ^{2}
{\mathcal {B}}^{-2}
\nonumber \\
& &
+ \alpha r^{4} \phi ^{4} \left( \frac {d\phi }{dr} \right) ^{2}
{\mathcal {A}}^{-1} {\mathcal {B}}^{-2}
+ \Lambda r^{4} \left( \frac {d\phi }{dr} \right) ^{2}
{\mathcal {A}}^{-1} {\mathcal {B}}^{-2}
\nonumber \\
& &
+ M^{2} \left( \frac {d\phi }{dr} \right) ^{2} {\mathcal {A}}^{-1}
{\mathcal {B}}^{-2}
+ \frac {1}{36} \frac {\Lambda }{\alpha } M^{2} \left( \frac {d\phi }{dr} \right) ^{2}
{\mathcal {A}}^{-1} {\mathcal {B}}^{-2}
\nonumber \\
& & \left.
+8\alpha r^{3} \phi ^{3} \frac {d\phi }{dr} {\mathcal {A}}^{-1}
{\mathcal {B}}^{-1}
+ \frac {4}{3} \Lambda r^{3} \phi \frac {d\phi }{dr}
{\mathcal {A}}^{-1} {\mathcal {B}}^{-1}
\right] ;
\label{eq:chargedpot}
\end{eqnarray}
with ${\mathcal {A}}$ and ${\mathcal {B}}$ as before (\ref{eq:AandB}).
As in the case of neutral black holes,
the Schr\"odinger equation (\ref{eq:sch}) with potential (\ref{eq:chargedpot})
could have been derived directly without using the conformal transformation (\ref{eq:cmet}).
However, our method has made the computations required significantly more tractable.

The potential ${\mathcal {C}}$ (\ref{eq:chargedpot}) reduces, in the limit
$\alpha \rightarrow - \Lambda /36$, to the potential ${\mathcal {U}}$ for the
neutral black holes.
This is as expected, since in this limit the charge $Q$ of the black hole tends to zero.

It is again straightforward to show that $\phi $, $d\phi /dr $
and ${\mathcal {B}}$ are all regular and non-zero everywhere
between the event and cosmological horizons.
However, unlike in the neutral case, for the charged MTZ
black holes it is no longer necessarily the case that ${\mathcal {A}}$
vanishes for some $r$ lying between $r_{+}$ and $r_{++}$.
Putting $\phi=\sqrt{6}$ at $r=r_{0}$ in (\ref{eq:scc}) and
using the inequalities $M<\frac{l}{4}$;
$r_{+}<r_{0}<r_{++}$ and $0<\frac{-\Lambda}{36\alpha}\leq 1$,
together with (\ref{eq:hor}), gives the
condition that ${\mathcal {A}}$ vanishes
somewhere between the event and cosmological horizons if
and only if
\begin{equation}
\label{eq:ps}
M<(-12\alpha)^{-\frac{1}{2}}\left(1+\frac{1}{6}\sqrt{\frac{-\Lambda}{\alpha}}
\right)^{-2}.
\end{equation}
When $Q=0$ we have from (\ref{eq:qm}) that
$\alpha=-\frac{\Lambda}{36}$ and (\ref{eq:ps}) becomes
$M<\frac{l}{4}$, which is always satisfied, retrieving the
result of section \ref{sec:pertneut}.
Similarly, in the limit $\Lambda \rightarrow 0$, the inequality
(\ref{eq:ps}) is trivially satisfied for all $M$, in accordance with
the results for the BBMB black hole~\cite{bronnikov}.

If equation (\ref{eq:ps}) is satisfied, then the second term in the potential
${\mathcal {C}}$ (\ref{eq:chargedpot}) has a double pole at $r=r_{0}$
(see Fig. \ref{fig:pot1} for a typical potential in this case).
Although there are other terms in ${\mathcal {C}}$ which have a single pole
at $r=r_{0}$, these will be sub-leading compared to the double pole.
In this case the behaviour of ${\mathcal {C}}$ near $r=r_{0}$ is
the same as for the potential ${\mathcal {U}}$~(\ref{eq:neutpot}):
\begin{equation*}
{\mathcal {C}} = -\frac {1}{4}
 \left[ N (r_{0}) \right] ^{2}
 \left( r - r_{0} \right) ^{-2}
+ O \left( r-r_{0} \right) ^{-1} .
\end{equation*}
As in the analysis of the neutral black holes, in this
case standard results in quantum mechanics \cite{bronnikov} allow us to
deduce that there are an infinite number of unstable modes.

\begin{figure}
\includegraphics[width=6cm,angle=270]{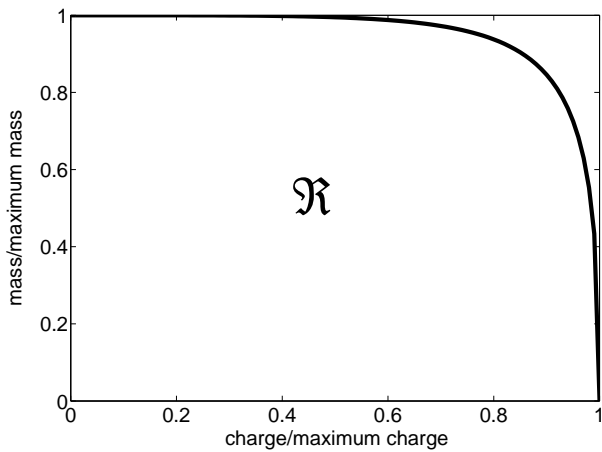}
\caption{The region $\Re $, described by Eq. (\ref{eq:Rdef}).
Inside $\Re $, the potential ${\mathcal {C}}$ has a pole
(see Fig. \ref{fig:pot1}),
and the corresponding black holes are unstable.
Outside $\Re $, the potential ${\mathcal {C}}$ is regular
(see Fig. \ref{fig:pot2}),
and numerical methods are necessary to show that the black holes
are unstable.}
\label{fig:region}
\end{figure}

It remains therefore to understand whether the inequality (\ref{eq:ps})
is satisfied by any or all of the charged MTZ black hole solutions.
We express this inequality in terms of the fractions of the
maximum possible values of $M$ ($l/4$) and $Q$ ($l{\sqrt {\pi }}/{\sqrt {2}}$) for any particular
value of $\Lambda$, and denote these model parameters by
$m$ and $q$ respectively.
The region $\Re $ of the $(m,q)$ phase space for which we have shown the
solution to be unstable is then
\begin{equation}
\Re =\left\{m(\Lambda),q(\Lambda),\Lambda :
m<\frac{4(1-q^{2})^{\frac{1}{2}}}
{(1+(1-q^{2})^{\frac{1}{2}})^{2}} \right\} .
\label{eq:Rdef}
\end{equation}
The region $\Re $ is shown in figure \ref{fig:region}, where it
can be seen that it makes up the vast majority of the available
parameter space.
Only black
holes carrying a very high charge, or those
nearly as large as
their universe (so that the event and
cosmological horizons nearly coincide),
are parameterized by variables $(m,q)$ lying outside the region $\Re $.
While it could be argued that such black holes are unphysical,
nevertheless for completeness we investigate the phase space
outside $\Re $.

\begin{figure}
\begin{center}
\includegraphics[width=6cm,angle=270]{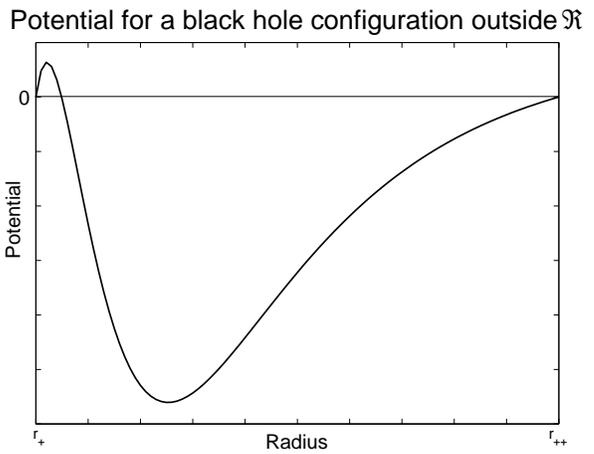}
\caption{Form of the potential ${\mathcal {C}}$ for
charged MTZ black hole configurations given by parameters $(m,q)$ lying outside the
region $\Re $.
The potential is regular everywhere between the event and cosmological horizons.
It is positive close to the event horizon, and then becomes negative.
As the values of $(m,q)$ approach the boundary of the region $\Re $, the
positive peak close to the event horizon becomes larger and larger.}
\label{fig:pot2}
\end{center}
\end{figure}

A typical potential for a point outside $\Re $ is shown in
Fig. \ref{fig:pot2}.
The potential is regular everywhere between the event and cosmological horizons,
where it vanishes.
There is a region close to the event horizon where the potential is positive;
elsewhere it is negative.
For black holes described by parameters $(m,q)$ lying close to the boundary of the region
$\Re $, the positive peak close to the event horizon becomes very large.

\begin{figure}
\begin{center}
\includegraphics[width=6cm,angle=270]{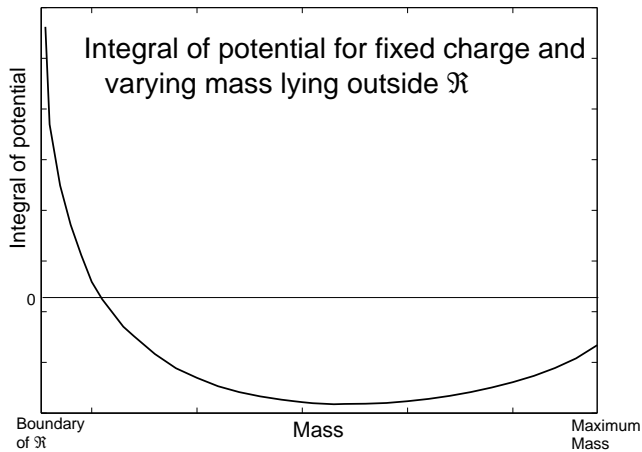}
\caption{The integral of the potential ${\mathcal {C}}$ at points
lying outside the region $\Re $.  In this case the charge is fixed to be 0.95 of the maximum charge,
and the values
of the mass vary from the value on the boundary of $\Re $ up to the maximum.}
\label{fig:integral}
\end{center}
\end{figure}

We show in the appendix that potentials of this form with
\begin{equation}
\int _{-\infty }^{\infty } {\mathcal {C}} \, dx_{*} <0  ;
\label{eq:intcond}
\end{equation}
must have at least one bound state, and therefore the corresponding
black holes will be unstable.
We have calculated numerically the integral on the left-hand-side
of (\ref{eq:intcond}), and some typical results are shown in
Fig. \ref{fig:integral}.
We find that for most configurations outside $\Re $,
the integral (\ref{eq:intcond}) is indeed negative, so that
those black holes are unstable.

However, for black hole configurations very close to the boundary
of $\Re $, we find that the integral becomes
positive (due to the large positive peak in the potential
close to the event horizon).
For such black holes their stability has to be investigated numerically,
each configuration being considered separately.
A simple way to check the existence of negative
eigenvalues of (\ref{eq:sch}), without actually finding the values of
these eigenvalues, is to apply continuity arguments of the type outlined in Ref. \cite{courant}.
If, for some $\sigma ^{2}<0$ (which is not necessarily an eigenvalue of (\ref{eq:sch})),
the solution of (\ref{eq:sch}) satisfying the boundary conditions at, say,
$x_{*} \rightarrow -\infty $ ($r=r_{+}$) has one zero before possibly diverging as
$x_{*} \rightarrow \infty $ ($r\rightarrow r_{++}$), then
there is at least one negative eigenvalue of (\ref{eq:sch}).
We studied various black holes
corresponding to points $(m,q)$ lying outside the region $\Re $ and
such that the integral on the left-hand-side of (\ref{eq:intcond}) is positive,
and found, in each case, a value of $\sigma ^{2}<0$ satisfying this condition.
Therefore, for all the cases we studied, there are
negative eigenvalues of (\ref{eq:sch}), showing instability.

\section{Conclusions}
\label{sec:conc}

We have investigated the stability of the de Sitter black hole solutions of
Einstein's equations due to Martinez et al \cite{mtz}.
The model contains a conformally coupled scalar field with a quartic
self-interaction potential and also an electromagnetic field in the
case that the black holes are charged.
We have shown by analytic methods that
all the neutral black holes are unstable,  as are
the charged black holes in the vast majority of the phase space.
Outside this region of phase space, we have used
numerical methods to show that the black holes
are unstable.

The analysis differs from that for the analogous, asymptotically flat
BBMB black holes, where the analytic approach suffices for both the neutral
and all the charged configurations.
However, the conclusions are the same, that the neutral and charged MTZ black holes,
like the BBMB black holes are unstable.
We are therefore able to complete the following table,
describing the behaviour of conformally coupled scalar field black hole hair (cf. that in Ref. \cite{ew}):
\newline
\begin{center}
\begin{tabular}{cc}
\hline
$\Lambda =0$  & Unstable hair
\\
$\Lambda >0$  & Unstable hair
\\
$\Lambda <0$  & Stable hair
\\
\hline
\end{tabular}
\end{center}

\begin{acknowledgments}
The work of E.W. is supported by the Nuffield Foundation, grant reference
number NUF-NAL 02, and that of P.M.Y. is supported by a studentship from the EPSRC.
We would like to thank C. van de Bruck for helpful discussions.
\end{acknowledgments}

\appendix*
\section{Instability if the integral of the potential is negative}
\label{sec:proof}

In this appendix we shall prove a result needed in the
stability analysis of the charged MTZ solutions in section \ref{sec:pertcharged};
namely that if
\begin{equation}
\label{eq:Win1}
\int^{\infty}_{-\infty}\mathcal{C} \,dx_{*}<0;
\end{equation}
then there is at least one negative eigenvalue of
(\ref{eq:sch}), and the corresponding
black hole configuration is unstable.

The starting point is a result from Ref. \cite{wald}:
if there exists a twice differentiable function $f$ such that
$f(-\infty)=f(\infty)=0$, and
\begin{equation}
\label{eq:Wald1}
\int^{\infty}_{-\infty}\left[\left(\frac{df}{dx_{*}}\right)^{2}+
\mathcal{C}f^{2}\right]dx_{*}<0 ;
\end{equation}
then there is at least one negative eigenvalue of
(\ref{eq:sch}).

In this appendix we show that the condition (\ref{eq:Wald1})
is satisfied if (\ref{eq:Win1}) holds.
The proof is straightforward, so we shall just briefly outline the
key steps.

We define a sequence of functions $f_{k}(x_{*}):k\in\mathbb{N}$, such
that each $f_{k}(x_{*})$ tends to zero as $|x_{*}|\rightarrow \infty $,
following \cite{galtsov}:
\begin{equation}
f_{k}(x_{*})=Z\left(\frac{x_{*}}{k}\right);
\label{eq:fdef}
\end{equation}
where
\begin{eqnarray*}
Z(u) & = & Z(-u) ; \\
Z(u) & = & 1\quad \textrm{for}  \quad u\in[0,p]; \\
Z(u) & = & 0\quad \textrm{for}\quad u>p+1 ;
\end{eqnarray*}
and
\begin{equation*}
\quad-q\leq Z'(u)\leq 0\quad \textrm{for}\quad u\in[p,p+1] ;
\end{equation*}
\smallskip
\newline
where $p$ and $q$ are positive constants.
Therefore $f_{k}(x_{*})$ is an even function
stretched horizontally for increasing $k$.

Now
$\frac{df_{k}}{dx_{*}}$ is non-zero only on the intervals
\begin{equation*}
I_{1}=[-k(p+1),-kp];
\qquad  I_{2}=[kp,k(p+1)] ;
\end{equation*}
and decreases linearly with $k$ for corresponding points as
$f_{k}$ is scaled.
Hence
$\left(\frac{df_{k}}{dx_{*}}\right)^{2}$ correspondingly
decreases as $\frac{1}{k^{2}}$, whereas the widths of the
intervals $I_{1}$ and $I_{2}$ increase linearly with
$k$.
It follows that
\begin{equation*}
\int^{\infty}_{-\infty}\left(\frac{df_{k}}{dx_{*}}\right)^{2}dx_{*} ;
\end{equation*}
decreases proportionally as $\frac{1}{k}$ as $k$ increases,
hence the first term in (\ref{eq:Wald1}) can be made arbitrarily small
for sufficiently large $k$.

We know that for black hole configurations corresponding to points lying
outside the region $\Re $
(defined in section \ref{sec:pertcharged}),
the potential ${\mathcal {C}}$ is continuous
and tends to zero as $x_{*}\rightarrow\pm\infty $, or as
$r\rightarrow r_{+}$ or $r_{++}$.
From this and the definition of $f_{k}(x_{*})$ (\ref{eq:fdef}) we deduce that
\begin{eqnarray*}
\left|\int^{\infty}_{-\infty}{\mathcal {C}} \,dx_{*}
-\int^{\infty}_{-\infty}\left(f_{k}(x_{*})\right)^{2}{\mathcal {C}} \, dx_{*}\right|
& &  \\
<\int_{-\infty}^{-pk}\left|{\mathcal {C}}\right|dx_{*}+
\int^{\infty}_{pk}\left|{\mathcal {C}}\right|dx_{*}. & &
\end{eqnarray*}
In addition, the potential
${\mathcal {C}}$ approaches zero linearly with
respect to $r$ very near $r_{+}$ and $r_{++}$, so
equation (\ref{eq:tort2}) gives us that for large
$\left|x_{*}\right|$, the potential $\left|{\mathcal {C}}\right| $ decreases
to zero exponentially with $\left|x_{*}\right| $.
Hence, both
$\int_{-\infty}^{-pk}\left|{\mathcal {C}}\right|dx_{*} $
and $\int^{\infty}_{pk}\left|{\mathcal {C}}\right|dx_{*}$ decrease
exponentially with increasing $k$.
Therefore,
$\int _{-\infty }^{\infty } {\mathcal {C}} \, dx_{*}$ will be
arbitrarily close to $\int _{-\infty }^{\infty } {\mathcal {C}} f_{k}^{2} \, dx_{*} $
for sufficiently large $k$.
This gives us our result, that if
$\int^{\infty}_{-\infty}{\mathcal {C}} dx_{*}$ is negative we can pick a
sufficiently large $k$ such that equation (\ref{eq:Wald1}) holds with
$f=f_{k}$, and we have shown the solution to be unstable.

\end{document}